\documentclass[useAMS,usenatbib,usegraphicx]{mn2e}

\newcommand{\Rs}{\ensuremath{R_{\odot}}}
\newcommand{\Ms}{\ensuremath{M_{\odot}}}
\newcommand{\eg}{{\it e.g.}}
\newcommand{\ie}{{\it i.e.}}
\newcommand{\viz}{{\it viz.}}
\newcommand{\simgt}%
        {\,\hbox{\lower0.6ex\hbox{$\sim$}\llap{\raise0.6ex\hbox{$>$}}}\,}
\newcommand{\simlt}%
        {\,\hbox{\lower0.6ex\hbox{$\sim$}\llap{\raise0.6ex\hbox{$<$}}}\,}

\title[Stellar-mass black holes in star clusters]
{Stellar-mass black holes in star clusters: implications for gravitational
wave radiation}

\author[S.~Banerjee, H.~Baumgardt and P.~Kroupa]
{Sambaran Banerjee$^{1,2}$\thanks{E-mail: sambaran@astro.uni-bonn.de (SB);
holger@astro.uni-bonn.de (BG); pavel@astro.uni-bonn.de (PK)},
Holger Baumgardt$^{1}$ and Pavel Kroupa$^{1}$\\
$^{1}$ Argelander-Institut f\"ur Astronomie, Auf dem H\"ugel 71, 53121, Bonn, Germany\\
$^{2}$ Alexander von Humboldt fellow}

\begin{document}

\date{Accepted --- --- ---. Received --- --- ---; in original from --- --- ---} 

\pagerange{\pageref{firstpage}--\pageref{lastpage}} \pubyear{2009}

\maketitle

\label{firstpage}

\begin{abstract}
We study the dynamics of stellar-mass black holes (BH) in star clusters
with particular attention to the formation of BH-BH binaries, which are
interesting as sources of gravitational waves (GW). In the present study, we examine
the properties of these BH-BH binaries through direct N-body simulations of
star clusters using the NBODY6 code on Graphical Processing Unit (GPU) platforms.
We perform simulations for star clusters with $\leq 10^5$ low-mass stars starting
from Plummer models with an initial population of BHs, varying the cluster-mass and BH-retention fraction.
Additionally, we do several calculations
of star clusters confined within a reflective boundary mimicking only the core of
a massive star cluster which can be performed much faster than the
corresponding full cluster integration. We find that stellar-mass BHs
with masses $\sim 10\Ms$ segregate rapidly ($\sim 100$ Myr timescale) into the
cluster core and form a dense sub-cluster of BHs within typically $0.2 - 0.5$ pc
radius. In such a sub-cluster, BH-BH binaries can be formed through 3-body encounters, the
rate of which can become substantial in dense enough BH-cores. While most
BH binaries are finally ejected from the cluster
by recoils received during super-elastic encounters with the single BHs,
few of them harden sufficiently so that they can merge via GW emission
within the cluster. We find that for clusters with $N \ga 5\times 10^4$, typically
1 - 2 BH-BH mergers occur per cluster within the first $\sim 4$ Gyr of cluster evolution.
Also for each of these clusters,
there are a few escaping BH binaries that can merge within a Hubble time, most of the
merger times being within a few Gyr. These results indicate that intermediate-age massive clusters
constitute the most important class of candidates for producing dynamical BH-BH mergers.
Old globular clusters cannot contribute significantly
to the present-day BH-BH merger rate since most of the mergers from them would have
occurred much earlier. On the other hand, young massive clusters with ages less that
50 Myr are too young to produce significant number of BH-BH mergers. 
We finally discuss the detection rate of BH-BH inspirals by the ``LIGO'' and ``Advanced LIGO''
GW detectors. Our results indicate that dynamical BH-BH binaries constitute the
dominant channel for BH-BH merger detection.
\end{abstract}


\section{Introduction}\label{intro}

Star clusters, \eg, globular clusters (henceforth GC), young and
intermediate-age massive clusters and open clusters
harbor a large overdensity of compact stellar remnants compared to that in
the field by virtue of their high density and the mass segregation
of the compact remnants towards the cluster core \citep{hv83}.
These compact stars, which are neutron stars (henceforth NS) and
black holes (henceforth BH), are produced by the stellar evolution of the most massive stars
within the first $\sim$ 50 Myr after cluster formation. These compact
stars, being generally heavier than the remaining low-mass stars of the cluster,
segregate quickly (within one or a few half-mass relaxation times) to the cluster core,
forming a dense sub-cluster of compact stars. Such a compact-star sub-cluster
is of broad interest as it efficiently produces compact-star binaries through
dynamical encounters \citep{hv83,h.et.al92}, \eg,
X-ray binaries and NS-NS and BH-BH binaries, which
are of interest for a wide range of physical phenomena. For example, X-ray
binaries are the primary sources of GC X-ray flux, while mergers of tight NS-NS binaries
(through emission of gravitational waves) is the most likely scenario
for the production of short duration GRBs and both NS-NS and BH-BH mergers are very important
sources of gravitational waves (henceforth GW) detectable by future gravitational
wave observatories like ``Advanced LIGO'' (AdLIGO) and ``LISA'' \citep{as2007}.
Double-NS systems are also interesting because they could be observable as double-pulsar 
systems \citep{rsm2008}, allowing important tests of General Relativity \citep{krm2008}. 
In the present work, we investigate the dynamics of stellar-mass BHs in star clusters,
with particular emphasis on dynamically formed BH-BH binaries.
Such binaries are strong sources of GWs as they spiral-in
through GW radiation, a process detectable out to several thousand 
Mpc distances.

Tight BH binaries that can merge within a Hubble time can also
be produced in the galactic field from tight stellar binaries,
which result from stellar evolution of the components
(\eg, \citealt{bul2003,osh2007,osh2008,bky2007}). 
However, \citet{bky2007} have shown with revised
binary-evolution models that the majority of potential BH-BH
binary progenitors actually merge because of common envelope (CE) evolution
which occurs when any of the binary members crosses the Hertzsprung gap.
They found that this reduces the merger rate by a factor as large as
$\sim 500$ and the resulting AdLIGO detection rate of BH-BH binary mergers
in the Universe from primordial binaries is only $\sim 2$ yr$^{-1}$,
subject to the uncertainties of the CE evolution model
that has been incorporated. Note that the corresponding NS-NS merger rate,
as the above authors predict, is considerably higher,
$\sim 20$ yr$^{-1}$. In view of the above result, the majority of
merging BH binaries are possibly those that are formed dynamically
in star clusters.

As studied by several authors earlier \citep{mt2006,mak2007},
black holes, formed through stellar evolution,
segregate into the cluster core within $\sim 0.3$ pc and form a sub-cluster of BHs,
where the density of BHs is large enough that BH-BH binary formation
through 3-body encounters becomes important \citep{hh2003}. These dynamically
formed BH binaries then ``harden'' through repeated super-elastic encounters
with the surrounding BHs \citep{h75,bg2006}.
The binding energy of the BH binaries released is
carried away by the single and binary BHs involved in the encounters. This
causes the BHs and the BH binaries to get ejected from the BH-core
to larger radii of the cluster and they heat the cluster while sinking
back to the core through dynamical friction \citep{mak2007}. Most of the energy
of the sinking BHs is deposited in the cluster core
as the stellar density is much higher there.
As the BH binaries harden, the encounter-driven recoil becomes stronger and finally
the recoil is large enough that the encountering single BH and/or the BH binary
escape from the cluster (see Sec.~\ref{res}, also \citealt{bcq2002}). Because of the associated mass-loss from
the cluster core, this also results in cluster heating. These heating mechanisms
result in an expansion of the cluster, as studied in detail by several authors,
\eg, \citet{mt2006,mak2007}. \citet{mak2007} found notable
agreement between the core expansion as obtained from their N-body
simulations with the observed age-core radius correlation for star clusters in
the Magellanic Clouds.

The dynamics of stellar mass BHs and formation of BH binaries in star clusters and
the resulting rate of GW emission from close enough BH binaries
has been studied by several authors using N-body integrations \citep{pzm2000},
numerical 3-body scattering experiments \citep{gul2004} or Monte-Carlo methods \citep{olr2006}.
\citet{pzm2000} considered the rate of mergers
of escaping BH binaries from various stellar systems,
\eg, massive GCs, young populous clusters and galactic nuclei.
They estimated the merger rates within 15 Gyr
(their adopted age of the Universe) by assuming
the binding-energy ($E_b$) distribution of the escaping
BH binaries to be uniformly distributed in $\log{E_b}$,
as inferred from simulations of $N \approx 2000$ or 4000 star clusters
(see \citealt{pzm2000} for details). Considering the space-densities of the
different kinds of star clusters,
they found that while for LIGO the corresponding total (\ie, contribution from
all types of star clusters) detection rate is negligible, it can
be as high as $\sim 1$ day$^{-1}$ for AdLIGO. \citet{gul2004}
performed sequential numerical integrations
of BH binary-single BH close encounters in a uniform stellar
background, where, in between successive encounters, the BH-binaries were evolved
due to GW emission. From such simulations, these authors studied the growth of BHs
through successive BH-binary mergers for the first time.
In a more self-consistent study of the growth of BHs and the possibility of formation of
intermediate mass black holes through successive BH-binary mergers,
\citet{olr2006} used the Monte-Carlo approach and considered ``pure'' BH clusters
that are dynamically detached from their parent clusters which can be expected to form
due to mass stratification instability \citep{spz}. These authors
considered dynamically formed BH binaries from 3-body encounters in such BH clusters
utilizing theoretical cross-sections of 3-body
binary formation (see \citealt{olr2006} and references therein).
Furthermore, they included both binary-single and binary-binary encounters.
Considering the sub-set of BH binaries formed
in the BH clusters that merge within a Hubble time, they determined typically a few
AdLIGO detections per year for old GCs.

While such results are remarkable and promising,
a more detailed study of the dynamics of stellar-mass BHs
in star clusters with a realistic number of stars is essential.
As the cross-sections of different processes governing the dynamics of BH binaries, \viz, 3-body binary
formation, binary-single star encounters and ionization have different
dependencies on the number of stars $N$ of the cluster, an extrapolation to much different
$N$ can be problematic. Hence it is important to consider clusters with 
values of $N$ appropriate for massive clusters or GCs. Also, exact treatments
of the various dynamical processes are crucial for realistic predictions of BH-BH merger
rates. Finally, a more careful study of the different dynamical processes leading
to the formation and evolution of BH binaries in star clusters is needed
to understand better under which conditions tight inspiralling BH binaries can be
formed dynamically from a star cluster. 

In the present work, we make a detailed study of
the dynamics of BH-BH binaries formed in a BH sub-cluster, as introduced above.
In particular, we investigate whether hard enough BH binaries
that can merge via gravitational radiation in a Hubble time
within the cluster or after getting ejected from the cluster, can be formed
in such a sub-cluster. To that end, we perform direct N-body integrations of concentrated
star clusters (half-mass radius $r_h \leq 1$ pc) consisting of 
$N \leq 10^5$ low-mass stars in which a certain number of stellar-mass BHs
is added, representing a star cluster with an evolved stellar population.

The present paper is organized as follows: In Sec.~\ref{sim} we describe
our simulations in detail. We discuss the various elements and assumptions
of the simulations and summarize all the runs that we perform (Sec.~\ref{runs}).
We also discuss the use of a reflective boundary to simulate only the core
of a star cluster (Sec.~\ref{reflct}). In Sec.~\ref{res} we discuss our
results with particular emphasis on the dynamical BH binaries formed during
the simulations. We discuss the BH-BH mergers that occur within the
clusters and also the merger timescales of the BH binaries that
escape from the clusters (Sec.~\ref{mrgesc}) and obtain the distributions
of merger-times for both cases (Sec.~\ref{tdist}). Finally, in
Sec.~\ref{discuss} we interpret our results in the context of
different types of star clusters that are observed and provide
estimates of BH-BH merger detection rates.

\section{Models}\label{sim}

\begin{figure}
\includegraphics[width=8.5cm,angle=0]{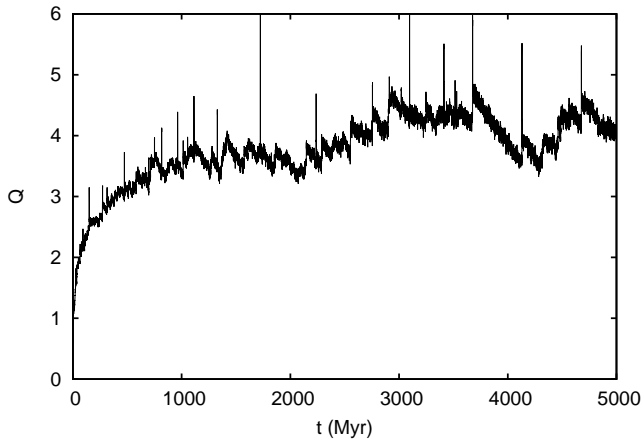}
\caption{Typical evolution of virial coefficient $Q$
of a star cluster with a reflective sphere. Shown is the evolution
of the run R3K180 of Table~\ref{tab1}. A rapid initial heating is observed,
caused by super-elastic encounters between BHs, followed by
a saturation of the heating curve, caused by the enhanced escape
rate of stars due to the heating (see text).}
\label{fig:qt}
\end{figure}

To study the dynamics of BHs in star clusters, we perform direct N-body
simulations using NBODY6 with star clusters in which a certain number of BHs are
added initially. The initial density distribution
follows a Plummer model with
half-mass radius $r_h \leq 1$ pc consisting of 
$N \leq 10^5$ low-mass main-sequence
stars in the mass-range $0.5\Ms \leq m \leq 1.0\Ms$.
Observed half mass radii for GCs and open clusters are usually larger,
but as we expect the clusters to expand considerably
due to the heating caused by the encounters in the BH core
(see Sec.~\ref{intro}), we begin with more concentrated clusters.

Black holes formed through stellar evolution are typically within the mass-range
$8\Ms \la M_{BH} \la 12\Ms$ for stellar progenitors
with solar-like metallicity \citep{cs2007,bky2009}.
The exact form of the BH mass-function is still debated
and depends on the metallicity of the parent stellar population
and the stellar wind mass-loss model \citep{bky2009}. In a dense stellar environment it
is further affected by the frequent stellar mergers and binary coalescence
\citep{bky2008}. In view of the uncertainty of
the BH mass-function, we consider only
equal-mass BHs, with $M_{BH} = 10\Ms$ in this work.
Such a value was also assumed by earlier authors, \eg, \citet{pzm2000},
\citet{bcq2002}.

A specified number of BHs are added to
each cluster, and we assume that the BHs follow the same spatial distribution
as the stars initially. The number of BHs added
depends on the BH retention fraction. We explore both full retention
and the case where half of the BHs are ejected from the cluster by
natal kicks. With such a cluster,
we mimic the epoch at which massive stars have already evolved and
produce BHs. While such a cluster is not completely representative of the
more realistic case in which the BHs are produced from stellar evolution
during the very early phases, it still serves the purpose of
studying the dynamics of the BHs since this
becomes important only later after segregation of the
BHs into the cluster core (see Sec.~\ref{res}).

We perform our simulations using NBODY6 code \citep{ar2003} enabled
for use with Graphical Processing Units (GPU). 
We have made arrangements in NBODY6 to put a specified number $N_{BH}$ of BHs 
of a given mass $M_{BH}$ by picking stars randomly from the cluster
and replacing them by the BHs.
Necessary rescalings have been done to compensate for the excess mass
gained by the cluster. These runs are performed on NVIDIA 9800 GX2/GTX 280/GTX 285
GPU platforms, located at the Argelander-Institut f\"ur Astronomie
(AIfA), University of Bonn, Germany, with the GPU enabled version
of NBODY6 (see \citealt{arhp}).

\subsection{BH-BH mergers}\label{bhmrg}

To evolve the BH-BH binaries due to GW emission, Peters'
formula \citep{pt64} is utilized in NBODY6 (also see \citealt{bm2006}),
which provides approximate orbit-averaged rates of evolution of binary
semi-major axis $a$ and eccentricity $e$ due to GW emission.
According to this formula, the merger time $T_{mrg}$ of an equal-mass BH-BH binary due to
GW emission is given by
\begin{equation}
T_{mrg} = 150{\rm Myr}\left(\frac{\Ms}{M_{BH}}\right)^3
\left(\frac{a}{\Rs}\right)^4(1-e^2)^{7/2}. 
\label{eq:tmrg}
\end{equation}

Peters' formula is limited up to the mass quadrapole terms of the radiating system.
Close to the merger, higher order PN terms become important, which
modifies the above value of the merger time \citep{bln2006}. In the present work however,
we are primarily interested in an overall statistics of the
merger events rather than the detailed orbit of a single BH-BH merger,
so that the higher order PN corrections are not crucial.
Hence we restrict ourselves to Peters' formula. Moreover, the orbit of a tight BH-BH
binary gets modified by dynamical encounters, which will anyway modify its
merger time to a much larger extent compared to that for the unperturbed
binary with the higher PN terms.

In NBODY6, the orbital evolution of compact binaries is also considered
when the binary is inside a hierarchy. Thus a tight BH binary will continue to
shrink even if it acquires an outer member forming a hierarchical triple,
which can often happen due to the strong focussing effect of BH binaries.
Also, energy removal due to GW (bursts) during a close hyperbolic passage
between two BHs is considered in the code.

Numerical simulations of BH-BH mergers (see \citealt{hg09} for an excellent review)
indicate that for unequal-mass BHs or even for equal-mass BHs with
unequal spins, the merged BH product acquires a velocity-kick of typically
100 km s$^{-1}$ or more due to an asymmetry in momentum outflow
from the system, associated with the GW emission.
Although we consider equal-mass BHs, the merger kicks associated
with the inequality of the spins of the merging BHs would be generally
sufficient to eject merged
BHs from the cluster. Therefore, in our simulations we provide an arbitrarily large kick of
150 km s$^{-1}$ immediately after a BH-BH merger,
to make sure that the merged BH escapes.

\subsection{Computations}\label{runs}

To study the rate of BH-BH mergers coming from a star cluster,
we perform simulations of isolated star clusters with single low-mass
stars and BHs as mentioned above.
The formation of the BH-core through mass segregation and its dynamics
remain largely unaffected by the presence of a tidal field, which
mainly affects stars near the tidal boundary.
While the enhanced removal of stars accelerates the core-collapse
of the cluster (see \eg, \citealt{spz}, Ch.~3), the latter is more strongly enhanced by the collapse
of the BHs themselves ($\sim 100$ Myr timescale, see below), so that
the effect of any tidal field is only second-order. Hence, isolated
clusters are good enough for our purposes. Further,
for simplicity, we do not take into account primordial binaries
in this initial study. The primordial binary fraction in GCs and their
period distribution is still widely debated \citep{a98, blz2002, sva2009}.
The presence of a primordial binary population can however significantly
influence the dynamics of stellar-mass BHs which we defer for a
future more detailed study.

For solar-like metallicity, Eggleton's stellar evolution model (\citet{eft89},
adapted in NBODY6) gives about $N_{BH} \approx 200$ BHs
for a cluster with $N=10^5$ stars following a Kroupa IMF \citep{krp2001}.
The above $N_{BH}$ (or its proportion with $N$) is thus an upper limit to the number of BHs
in a GC that corresponds to a full retention (\ie, no or low natal kicks for
all BHs).

We perform 2 simulations with $N = 4.5 \times 10^4$ and
$N_{BH}=80$, 2 runs with $N = 6.5\times 10^4$, $N_{BH}=110$, \ie,
about full BH-retention. Two of the above runs are repeated with half
the above $N_{BH}$s. Also, one run with $N = 5\times 10^4$
and excess $N_{BH}=200$, appropriate for a top-heavy IMF has been
performed. Finally, we do 2 runs with $N = 10^5$ with $N_{BH} = 80$ 
(about 50\% retention fraction) and $200$ (full retention). All the clusters consist
of low-mass stars between $0.5\Ms \la m \la 1.0\Ms$ with a Kroupa
IMF and the BHs have $M_{BH} = 10\Ms$, as discussed in detail
in the beginning of the section.

In addition to these systems, we also study stellar-mass BHs in clusters with
smaller $N$ representing open clusters,
in order to estimate the lower-limit in
cluster mass for the occurrence of BH-BH mergers. We perform
10 runs with $N = 5\times10^3$, $N = 10^4$ and $N = 2.5 \times 10^4$ each
with full BH-retention (\ie, $N_{BH}=12$, $20$ and $50$ respectively).
Results of all our runs are summarized in Table~\ref{tab1}.

\subsection{Simulation of a GC core: reflective boundary}\label{reflct}

\begin{figure}
\includegraphics[width=8.5cm,angle=0]{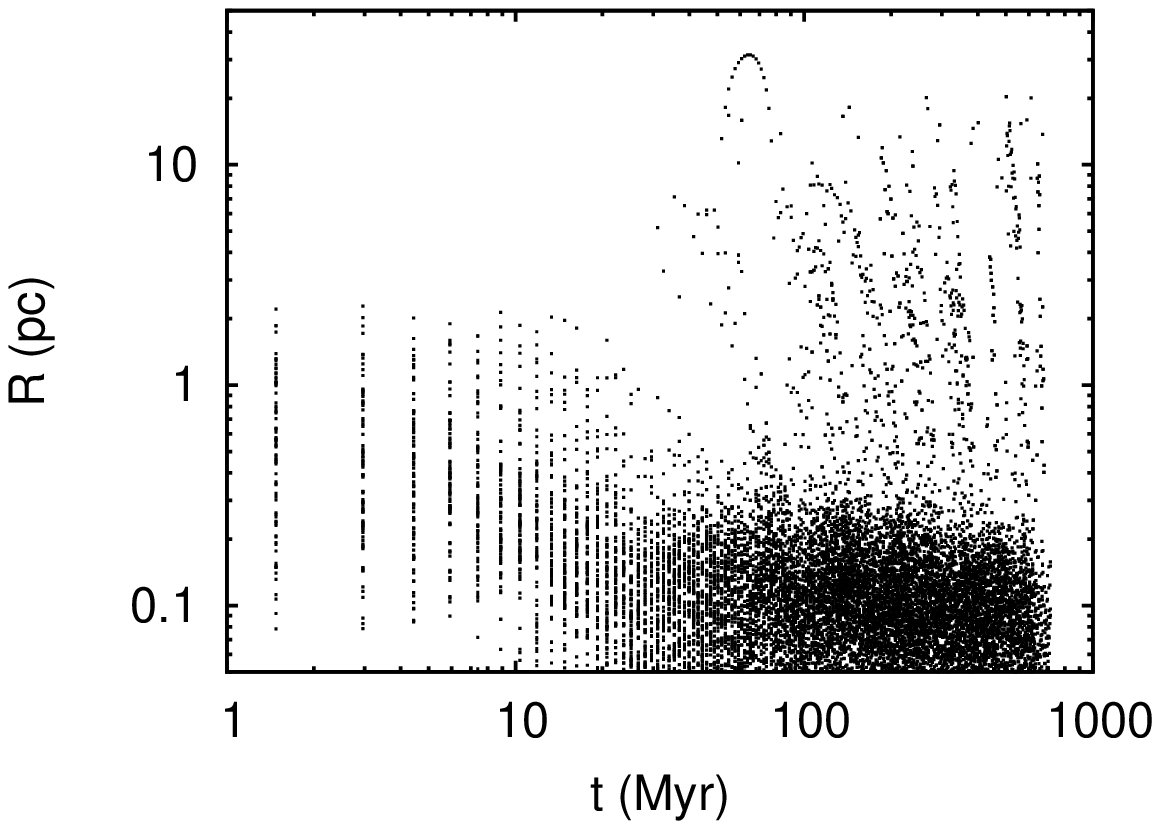}
\includegraphics[width=8.5cm,angle=0]{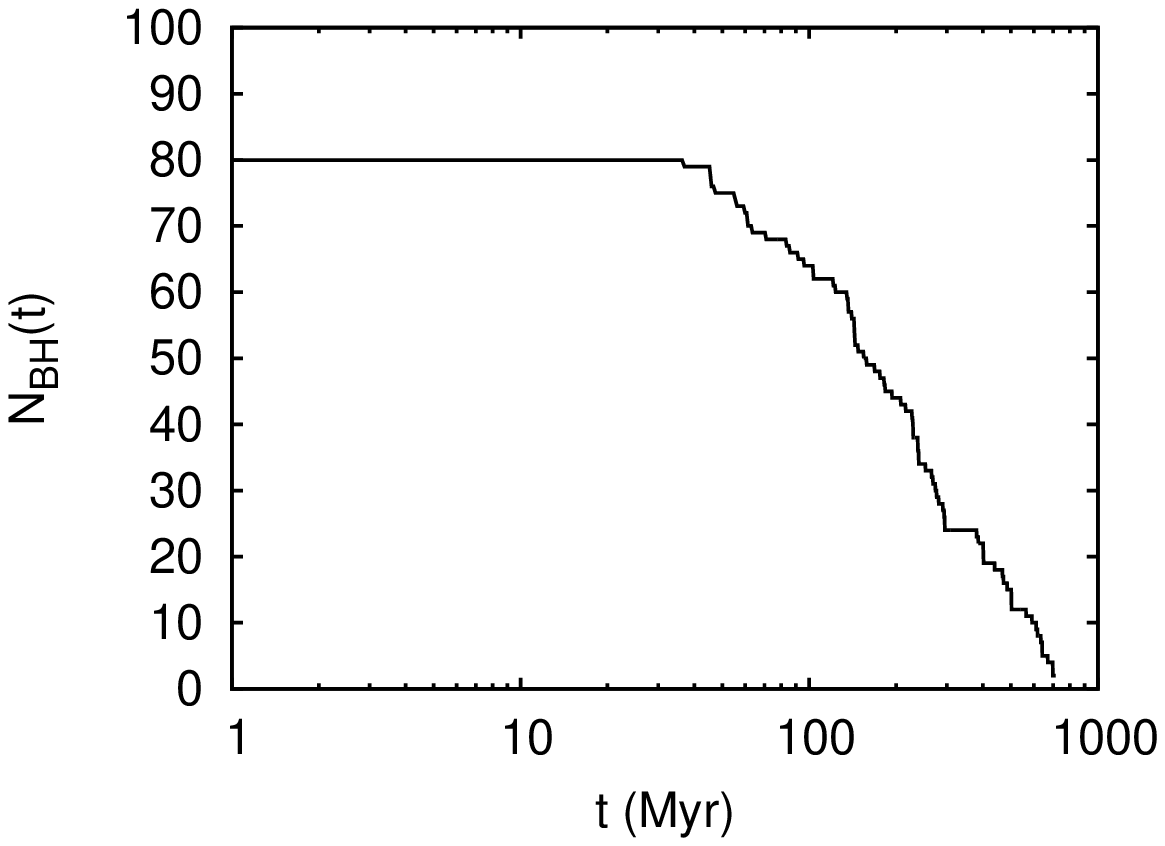}
\caption{Typical example of the mass segregation of BHs. Shown is
the radial position $R$ vs. time $t$ (top panel) of all BHs for model C50K80
of Table~\ref{tab1}. Each of the points represents a BH. The BHs segregate within 50 Myr during
which $N_{BH}$ remains unchanged (bottom panel).
As the BH sub-cluster becomes dense enough that BH-BH binary formation
takes place through 3-body encounters, BHs and BH binaries are ejected
from the BH-core and $N_{BH}$ starts to decrease.}
\label{fig:bhpos}
\end{figure}

We also perform simulations with a smaller number of stars and BHs that are confined within
a reflecting spherical boundary. With such a dynamical system
one can mimic the core of a massive cluster, where the BHs are concentrated
after mass segregation. The advantage of this approach is that one can simulate
the evolution of a massive cluster with much fewer stars. We simulate
$N = 3000 - 4000$ stars packed within $0.4$ pc. This provides
a stellar density of $\sim 10^4 \Ms {\rm ~pc}^{-3}$, appropriate for the
core-density of a massive cluster. The initial BH population
is taken to be $N_{BH} \approx 100$ or $200$, representing half or full
BH-retention respectively, of a $N = 10^5$ star cluster.
In this way, the simulation of the BH-core of a massive GC can be performed
much faster than the simulation of the whole cluster.

To implement the reflective boundary, we simply consider all outgoing stars
beyond the reflective sphere $R_s$ within each block and reflect them elastically.
This is done in the main integration loop.
For these stars, we compute the force polynomials \citep{ar2003} separately
(using the CPU) which makes the code slower by factors of 2-3
depending on the stellar density. While the behavior of a N-body code
in presence of a reflective boundary needs to be studied in more detail, the
present implementation seems to be fairly stable with single-stars and BHs,
with energy errors marginally larger than that for a free cluster.
(In fact, in the NBODY6 package, some routines are already available for optionally
implementing the reflective boundary). For stars and BHs
beyond a pre-assigned speed $v_{esc}$, representing the escape speed of the host cluster,
we do allow them to escape through the reflective boundary.

\begin{figure*}
\includegraphics[width=13.0cm, height=8.5cm, angle=0]{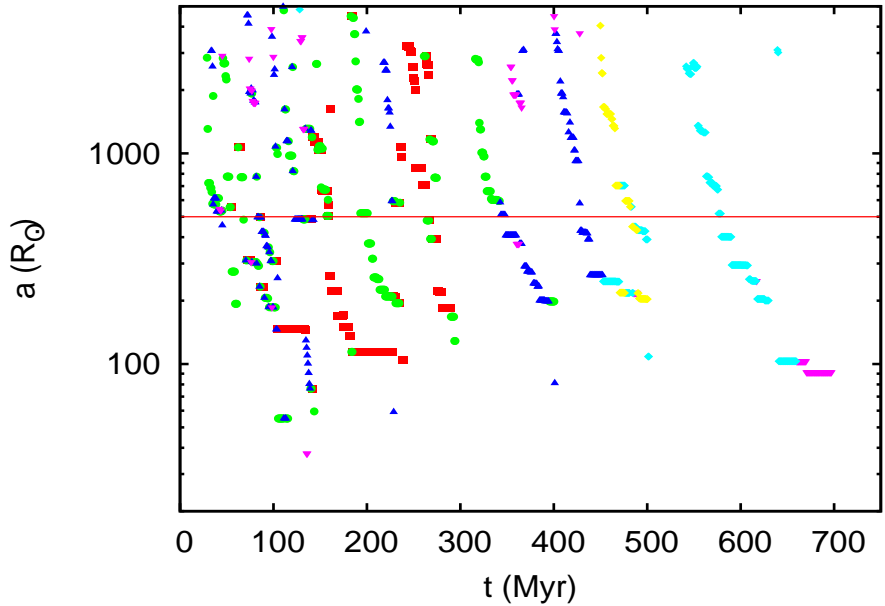}
\caption{Semi-major-axis $a$ vs. evolution time $t$ for all
BH-BH pairs formed for the model C50K80, where a different symbol is
used for each newly formed BH-BH binary. The sequences of the points
indicate that each BH binary initially hardens
quickly to smaller values of $a$ where the hardening rate decreases.
The estimated upper-limit $a_{esc}$ for which a BH-BH pair can escape due to recoil
from an encounter with a single BH is indicated by the horizontal line.
Hardening rates of all BH pairs terminate below this line.
Note that different traces of the same symbol represent different
hardening sequences following encounters.}
\label{fig:bhard}
\end{figure*}

It is of course important to note that a cluster simulation within a reflective
boundary cannot represent the evolution of a free cluster, in particular,
since the cluster expands due to the BH-heating which continually reduces the stellar
density in the core. With the reflective boundary, the stellar density still
decreases because of the escape of stars with speed $v > v_{esc}$,
but it generally does not mimic the cluster expansion.
The cooling effect of these escaping stars also helps to
inhibit the runaway heating of the cluster due to the binding
energy released by binary-star/binary-binary encounters.
The escaping single BHs and BH binaries are particularly efficient in this mass-loss
cooling due to their larger masses. Fig.~\ref{fig:qt} shows a typical example
from one of our runs (R3K180 of Table~\ref{tab1}, see below) in which the behavior
of the virial coefficient $Q\equiv T/V$ ($T=$ total kinetic energy of the system, $V=$ total
potential energy) is shown. While $Q$ initially increases
quickly indicating rapid heating of the cluster, the escape rate of stars and BHs
also increases due to the increased velocity dispersion. The latter
effect in turn increases the cooling rate so that $Q$ finally tends to flatten
to a value for which the dynamical heating is balanced by the escape-cooling
of the cluster. For the reflective boundary clusters used in our simulations
(see above), the initial velocity-dispersion
of the stars within the reflective boundary is chosen to be
$\overline v_0\approx 7$ km s$^{-1}$ and escape speed $v_{esc} \approx 24$ km s$^{-1}$.
Typically, during each run, the virial coefficient $Q$ increases by a
factor of about 6.5 and most of the growth takes place during the
first few hundred Myr, as in Fig.~\ref{fig:qt}.
As $\overline v \sim Q^{1/2}$, it increases to
$\approx 17$ km s$^{-1}$, which is typical for the central velocity
dispersion of a massive cluster.

In a reflective boundary cluster,
the BH binaries start forming immediately after the start of the integration since
the system already initiates with a BH-core.
We perform a set of 6 runs with reflective boundary clusters (see Table~\ref{tab1}). 
We shall discuss results of our reflective boundary simulations in Sec.~\ref{res}.

\section{Results: Black hole binary mergers and escapers}\label{res}

In this section we discuss the results of the simulations introduced in
Sec.~\ref{runs}. As already mentioned, we focus on tight BH-BH binaries both within
the cluster and the escaped ones that can merge within a few Gyr.
Table~\ref{tab1} provides an overview of the results of our simulations
for both isolated and reflective clusters.
It shows the number of BH-BH mergers
within the cluster for each computation and the
corresponding merger times. Also, for each computation,
the numbers of escaped BH-BH binaries that merge within 3 Gyr are
shown. For convenience of discussion, we utilize one of these models, \viz, 
the one with identity C50K80 (see Table~\ref{tab1}) ---
all others generally posses similar properties.

Fig.~\ref{fig:bhpos} (top panel) demonstrates the mass segregation of the BHs in the cluster which
takes about 50 Myr. Before the BHs segregate to within about $0.3 {\rm~pc}$
of the cluster center, the BH-density is small so dynamical encounters
among the BHs are not significant and $N_{BH}$ remains constant. As the
BHs segregate within about $0.3 {\rm~pc}$ of the cluster core, the BH-density
of this sub-cluster becomes high enough to form BH-BH binaries through 3-body encounters.
Once BH-BH binaries start forming, single and BH binaries begin to escape from the
BH-core by the recoils received due to repeated super-elastic encounters between
the single and binary BHs (see Sec.~\ref{intro}). In Fig.~\ref{fig:bhpos} (top panel) one can
clearly distinguish the two phases of the BH subsystem --- the initial segregation
phase and the BH-core formation, the radial positions of the BHs in the latter
phase being scattered outwards due to the recoils. The decrease of $N_{BH}$ during this
phase is also shown in Fig.~\ref{fig:bhpos} (bottom panel).

The time $t_{\rm seg}$ taken by the BHs to segregate to the cluster core to form
the BH sub-cluster is essentially the core-collapse time for the initial BH cluster.
$t_{seg}$ is given by (see \eg, \citet{spz}, Ch.~3):
\begin{equation}
t_{\rm seg} = \frac{\langle m \rangle}{M_{BH}}t_{cc},
\label{eq:tseg}
\end{equation}
where, $t_{cc}$ is the core-collapse time of the host star cluster
itself (\ie, without the BHs) and $\langle m \rangle$ is its mean stellar mass.
According to numerical experiments by several authors (\eg, \citet{bg2002}),
a Plummer cluster takes about $t_{cc} \approx 15t_{rh}(0)$ to reach the core-collapse stage,
where $t_{rh}(0)$ is the initial half-mass relaxation time of the Plummer
cluster. For the above example,
$t_{rh}(0) \approx 73 {\rm ~Myr}$ (see Eqn.~(2-63) of \citet{spz})
and $\langle m \rangle\approx 0.6$ which gives $t_{seg} \approx 65 {\rm ~Myr}$
from Eqn.~(\ref{eq:tseg}). This roughly agrees with the formation time
of the central dense BH-cluster as observed in the 
N-body integration (see Fig.~\ref{fig:bhpos}).

\subsection{Dynamical black hole-black hole binaries}\label{bhbin}

A particular dynamically formed BH-BH binary can initially be very wide,
sometimes with semi-major-axis $a$ more than $5000\Rs$. However, it
then shrinks very rapidly due to repeated
super-elastic encounters with the surrounding BHs.
The collisional hardening rate is given by $\dot a_C \propto a^2$ (see \citealt{bg2006}
and references therein). The variation
in $a$ and $e$ is random due to the stochastic nature of the
encounters, but the binary hardens on average (see \eg, \citealt{hh2003},
Ch.~19 \& Ch.~21) and the overall hardening rate decreases with $a$.
These encounters include both flybys and exchanges with single BHs.
When the binary is wide, and the recoils it receives are low, so that
it usually remains in the cluster. However, as it becomes harder,
the recoils become stronger, and the binary is ejected
from the BH-core, but returns to the BH-core due to
dynamical friction (see Sec.~\ref{intro}). Finally, the recoil
becomes strong enough that the BH binary escapes from the cluster
(see also \citealt{bcq2002}). 

For equal mass binary-single star encounters, 40\% of the binding energy of the binary
is released per close encounter on average (\citealt{spz}, Eqn.~(6-25); \citealt{H.et.al92}).
This leads to the following expression for the average recoil velocity
of a BH binary due to encounters with single BHs:
\begin{equation}
\langle v_{rec}^2 \rangle = 6.75\times 10^{-2}
\frac{GM_{BH}}{a}.
\label{eq:vrec}
\end{equation}
Hence, the BH binary can be ejected due to recoil from
a cluster with escape velocity $v_{esc}$ for $a < a_{esc}$,
where $a_{esc}$ is obtained by setting the left hand side
of Eqn.~(\ref{eq:vrec}) to $v_{esc}^2$. For the above
Plummer cluster, we then get from the value of its
central potential (see \citealt{hh2003}, Table 8.1)  $a_{esc} \approx 500\Rs$,
which is shown as solid line in Fig.~\ref{fig:bhard}.
The escaped binaries in our calculations are generally found
to be harder than $a_{esc}$ except for a few systems, which escape
within triple-BHs.

Fig.~\ref{fig:bhard} provides an overall impression of the hardening of the BH binaries
where the values of $a$ for all BH-BH binaries formed during the computation are plotted.
Each newly formed BH pair is represented by a different symbol.
From the sequences of points, it can be seen
that each of the BH-BH pairs exhibits an overall tendency
of rapid hardening and the steep $a(t)$ dependence tends to flatten
as $a$ decreases. The termination of a particular curve generally indicates
escape of the corresponding BH binary from the cluster (or more rarely a
merger, see Sec.~\ref{mrgesc}). Typically, a BH binary leaves the cluster with
$a<200\Rs$ (see Fig.~\ref{fig:bhesc}).

\subsection{Mergers and escapers}\label{mrgesc}

\begin{figure*}
\vspace{-2.5 cm}
\includegraphics[width=17.5cm,angle=0]{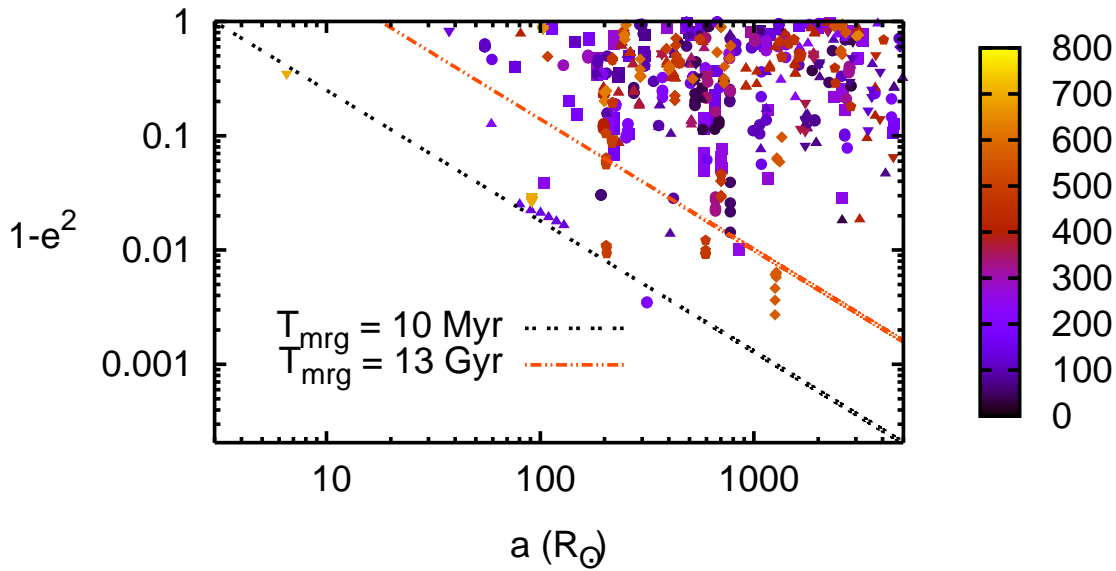}
\vspace{-1.3 cm}
\caption{Positions of all the BH binaries within the cluster in a $1-e^2$ vs. $a$
plane where different symbols are used to distinguish between
different BH pairs (for C50K80). The color-coding of the points, as indicated
by the color-scale, represents the evolution time (in Myr) of the cluster
at which they appear at a particular location in the
above plane. See text for details.}
\label{fig:aeplot}
\end{figure*}

Can a BH-pair be hardened to small enough $a$
(and eccentric enough at the same time) for it to merge via GW radiation?
To investigate this question, we consider the positions
of the BH-BH binaries within the cluster in a $a {\rm~ vs.~} (1-e^2)$ plane (Fig.~\ref{fig:aeplot}),
where each newly formed BH pair is represented by a different symbol as in Fig.~\ref{fig:bhard}.
The points are color-coded with
the evolution time in Myr, the color-scale being displayed
on the right. Although for each BH-pair $a$ and $e$
fluctuate over the plane, the changes occur
on a collision time-scale which is $\sim$ Myr. Since the orbital
period of the binaries corresponding to these points is much shorter
(from $\sim$ days to years), these points generally represent
binaries which are stable over many orbits.
Overplotted on Fig.~\ref{fig:aeplot} are lines of
constant GW merger time, $T_{mrg}$, as given by Eqn.~(\ref{eq:tmrg}).
While most of the points have very large merger time, a few of
them do lie around the $T_{mrg}=10 {\rm ~Myr}$ line. This indicates that
these binaries are indeed hardened up to small enough $a$ and/or acquire sufficient
eccentricity that if they are left unperturbed, they can merge via GW emission
within several Myr. This feature
is found to be generally true for all the computations reported here (see Table~\ref{tab1}),
ie, each of them does produce BH-pairs that are capable of merging within
several Myr. It is important to note that the short merger timescale of
these binaries is due to their high eccentricities since they are generally
far too wide ($\sim 100\Rs$) to merge unless they are highly eccentric
(see Eqn.~(\ref{eq:tmrg})). This is also true for the escaping BH binaries
(see below).

However, these merging BH-pairs, particularly due to their large
eccentricity and focussing effect, can still be perturbed by further
encounters on timescales $\sim$ Myr which can
often prevent them from merging. In the particular example shown in
Fig.~\ref{fig:aeplot}, only one of the ``merging candidates''
($T_{mrg}\leq 10 {\rm ~Myr}$) could actually merge within the cluster (inverted triangles). For all the
other candidates, including the one which escaped (filled dots), the
eccentricity became much smaller due to encounters so that the merger timescales
increased considerably.

\begin{figure}
\includegraphics[width=9.2cm,angle=0]{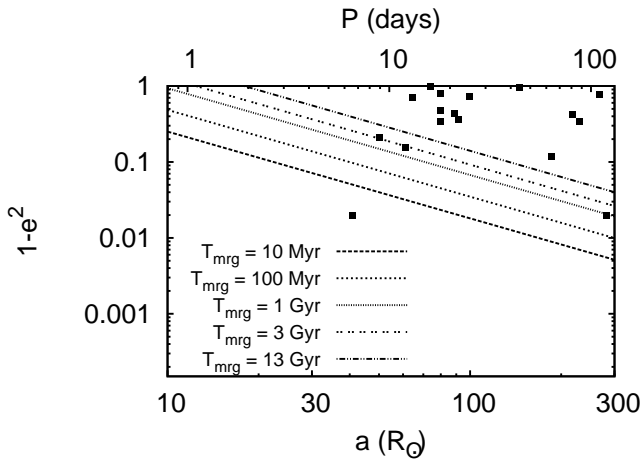}
\caption{Positions of all escaped BH binaries in a $1-e^2$ vs. $a$
plane for the model C65K110. Lines of constant merger times are
plotted as in Fig.~\ref{fig:aeplot}.}
\label{fig:bhesc}
\end{figure}

As for the escapers, since they remain unperturbed afterwards, all with
GW merger times smaller than a Hubble time are of interest.
However, we find that typically
for each cluster there are relatively few BH binaries with
GW merger times $T_{mrg} \ga 3{\rm ~Gyr}$ among those that merge
within a Hubble time (see \eg, Fig.~\ref{fig:bhesc} and Fig.~\ref{fig:mrgdist}).
In the above example, \ie, model C50K80, there is one BH-BH escaper with
$T_{mrg} \sim 1 {\rm ~Gyr}$.
A more interesting example is shown in Fig.~\ref{fig:bhesc} corresponding to the
model C65K110 of Table~\ref{tab1} where 2 binaries merge in $\approx 3$ Gyr,
one in $\approx 1$ Gyr and one in $< 10$ Myr.

The results of the present computations, as summarized in Table~\ref{tab1},
indicate that a medium-mass to massive cluster with sufficient BH-retention
is likely to have at least one BH-pair that can merge within the cluster within a time-span
of a few Gyr. Each of these clusters also typically eject a few
BH binaries that can merge within a Hubble time. In one of the
models (C50K200), one BH-BH hyperbolic collision also
occurred at $t_{mrg}\approx 820$ Myr. As such hyperbolic mergers did not show up
in any of the other models, they must be much rarer than GW driven mergers.

\begin{table*}
\centering
\caption{Summary of the computations performed for isolated clusters and those
with reflective boundary, see Sec.~\ref{runs} \& Sec.~\ref{reflct}.
The meaning of different columns is as follows:
Col.~(1): Identity of the particular model --- similar values with different names (ending with
A,B etc) imply computations repeated with different random seeds, Col.~(2): Total number of stars $N$,
Col.~(3): Number of simulations $N_{sim}$ with the particular cluster,
Col.~(4): Initial half-mass radius of the cluster $r_h(0)$ (isolated cluster) or radius of reflective sphere $R_s$,
Col.~(5): Initial number of BHs $N_{BH}(0)$, Col.~(6): Total number of BH-BH binary mergers within the
cluster $N_{mrg}$, Col.~(7): The times $t_{mrg}$ corresponding to the mergers, Col.~(8): Number
of escaped BH-pairs $N_{esc}$ --- the three values of $N_{esc}$ are those with
$T_{mrg} \la $ 3 Gyr, 1 Gyr and 100 Myr respectively, Col.~(9): BH-BH Merger rate  ${\mathcal R}_{\rm AdLIGO}$ detected by AdLIGO
assuming that the corresponding model cluster has a space density of $\rho_{cl}=3.5{\rm~}h^3$ Mpc$^{-3}$
(see Sec.~\ref{rate}).}
\begin{tabular}{lllclllcl}
\hline
Model name &    $N$   &  $N_{sim}$ & $r_h(0)$ or $R_s$ (pc)&  $N_{BH}(0)$ &    $N_{mrg}$  &  $t_{mrg}$ (Myr)  &  $N_{esc}$  & ${\mathcal R}_{\rm AdLIGO}$\\
\hline   
\multicolumn{8}{c}{Isolated clusters}\\
\hline
C5K12      & 5000    &   10       &   1.0              &   12         &       0        &  --- ---           & --- ---   &   --- ---	\\
C10K20     & 10000   &   10       &   1.0              &   20         &       0        &  --- ---           & --- ---   &   --- ---	\\
C25K50     & 25000   &   10       &   1.0              &   50         &       0        &  --- ---           &  3 1 1    &   --- ---	\\
C50K80     & 45000   &   1        &   1.0              &   80         &        1       &   698.3            &  3 1 0    & $28(\pm 14)$	\\
C50K80.1   & 45000   &   1        &   0.5              &   80         &        2       &   217.1, 236.6     &  3 2 1    & $35(\pm 15)$	\\ 
C50K40.1   & 45000   &   1        &   0.5              &   40         &        0       &   --- ---          &  1 1 1    & $7(\pm 7)$	\\  
C50K200    & 50000   &   1        &   1.0              &   200        &       2        &   100.8, 467.8     &  0 0 0    & $14(\pm 10)$	\\
C65K110    & 65000   &   1        &   1.0              &   110        &       1        &   314.6            &  4 2 1    & $35(\pm 15)$	\\
C65K110.1  & 65000   &   1        &   0.5              &   110        &       0        &  --- ---           &  4 3 1    & $28(\pm 14)$	\\
C65K55.1   & 65000   &   1        &   0.5              &   55         &       1        &   160.5            &  1 0 0    & $14(\pm 10)$	\\
C100K80    & 100000  &   1        &   1.0              &   80         &       2        &   219.4, 603.2     &  5 2 1    & $42(\pm 15)$	\\
C100K200   & 100000  &   1        &   1.0              &   200        &       0        &  --- ---           &  5 4 4    & $28(\pm 14)$	\\
\hline
\multicolumn{8}{c}{Reflective boundary}\\
\hline
R3K180     & 3000   &    1        &   0.4              &  180         &       1        &    1723.9          &  5 3 1    & $35(\pm 15)$	\\
R4K180A    & 4000   &    1        &   0.4              &  180         &       1        &    3008.8          &  2 2 1    & $21(\pm 12)$	\\
R4K180B    & 4000   &    1        &   0.4              &  180         &       2        &  100.2, 1966.5     &  2 1 0    & $28(\pm 14)$	\\
R3K100     & 3000   &    1        &   0.4              &  100         &       2        &  3052.8, 3645.9    &  1 1 0    & $18(\pm 10)$	\\
R4K100A    & 4000   &    1        &   0.4              &  100         &       2        &  104.4, 814.2      &  3 3 1    & $28(\pm 14)$	\\
R4K100B    & 4000   &    1        &   0.4              &  100         &       1        &    1135.3          &  3 3 3    & $28(\pm 14)$	\\
\hline
\end{tabular}
\label{tab1}
\end{table*}

\subsection{Merger-time distribution}\label{tdist}

\begin{figure}
\includegraphics[width=8.5cm,angle=0]{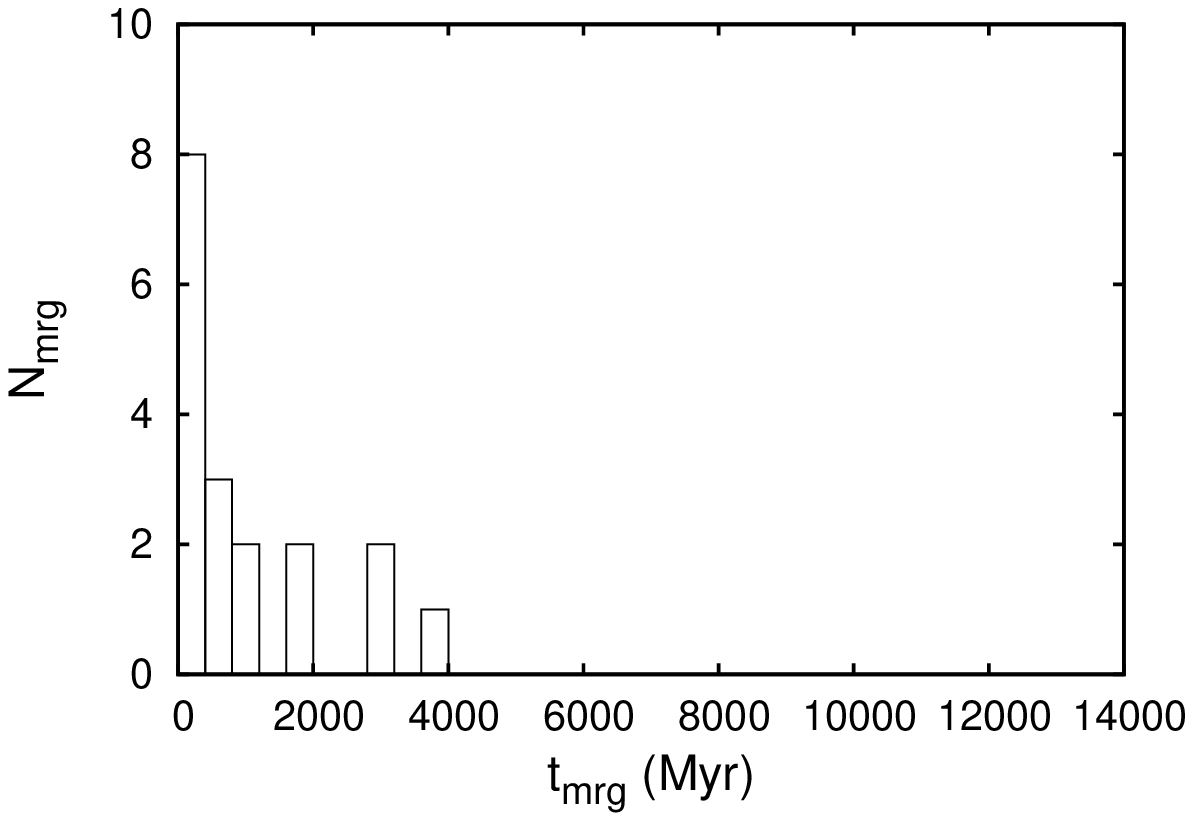}
\includegraphics[width=8.5cm,angle=0]{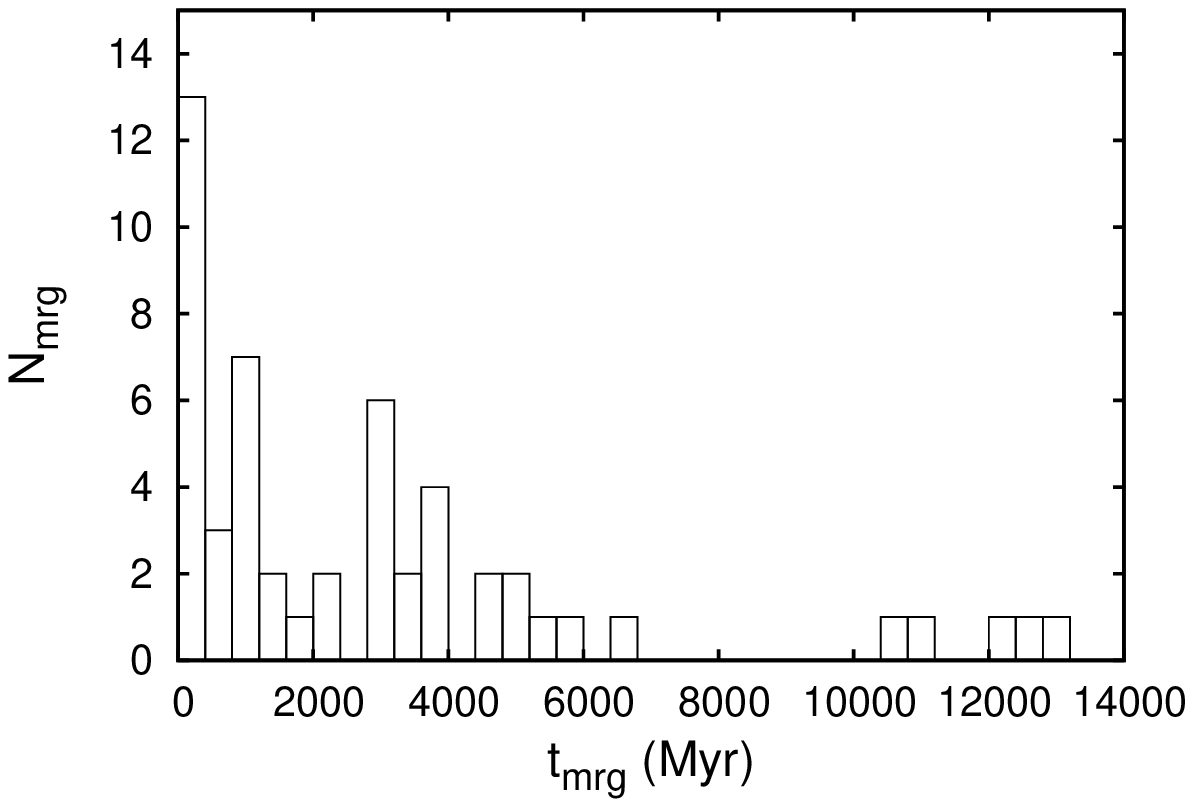}
\caption{{\bf Top:} Distribution of the merger times $t_{mrg}$ for BH binary mergers
within the cluster for the models of Table~\ref{tab1}. {\bf Bottom:} Distribution of
the merger times $t_{mrg}$ for escaped BH binaries
for the models of Table~\ref{tab1} (see text).}
\label{fig:mrgdist}
\end{figure}

As a representation of the merger time distribution of mergers within the cluster, a combined distribution of
$t_{mrg}$ of all the BH binaries that merged within the clusters is shown in Fig.~\ref{fig:mrgdist} (top panel). 
This is justified, as the values of BH-BH binary merger times $t_{mrg}$ within the cluster, shown in
Table~\ref{tab1}, apparently do not indicate any correlation of $t_{mrg}$ with the cluster parameters,
in particular the number of stars $N$.
Both the isolated clusters and the reflective boundary
clusters are included. The $t_{mrg}$ distribution in Fig.~\ref{fig:mrgdist} indicates that $N_{mrg}$ decreases
with $t_{mrg}$. This can be expected, as at later times, $N_{BH}$ in a cluster decreases,
so the hardening rate of BH binaries also decreases, making their merger less
probable. Comparing our results for clusters
with different values of $N$, we find that the rate of depletion of BHs from the
BH core during the first few Gyr does not depend appreciably on $N$.
In our computations, BHs are depleted within a few Gyr for most of the models.
Clusters with large $N_{BH}$s, \eg, C100K200, C50K200 do indicate longer BH-retention,
but this is due to the expansion of the cluster (and therefore the BH sub-cluster itself)
because of the heating by the BH-core (see Sec.~\ref{intro}) and hence it is unlikely
that tight BH-BH binaries can be formed during the later phases.

Fig.~\ref{fig:mrgdist} (bottom panel) shows the distribution of the
merger times for the escaping BH binaries from all the computations of Table~\ref{tab1}.
Note that the number of tight escapers in Table~\ref{tab1} also does not indicate
any dependence on $N$. For each escaper $t_{mrg}$ includes its time of escape $t_{esc}$, \ie,
$t_{mrg} = t_{esc} + T_{mrg}$, where $T_{mrg}$ is calculated from Eqn.~(\ref{eq:tmrg})
using the values of $a$ and $e$ with which it escaped.
The merging rate among escaped binaries also shows a decrease with time.
Such a decrease is expected from Eqn.~(\ref{eq:tmrg}). For example, if
we consider that the escapers that merge within a Hubble time have a mean
radius $\overline a \sim 50\Rs$ and a thermal eccentricity distribution
$dN_{mrg}/de \propto 2e$, it is easy to show from Eqn.~(\ref{eq:tmrg})
that $dN_{mrg}/dT_{mrg} \propto T_{mrg}^{-5/7}$.
The above merger-time distribution indicate that significantly more BH-BH mergers occur
outside the clusters, \ie, among the escapers, than within the clusters.
We shall return to this point in Sec.~\ref{discuss}. The median merger time for BH-mergers within
the clusters is $\overline t_{mrg} \approx 1 {\rm ~Gyr}$ and for the escaped BH binaries it is
$\approx 3 {\rm ~Gyr}$.

\section{Discussion}\label{discuss}

Our present study indicates that centrally concentrated star clusters,
with $N \ga 4.5\times 10^4$ are capable of dynamically producing
BH binaries that can merge within a few Gyr, provided a significant
number of BHs are retained in the clusters after their birth. The
results of our simulations (see Table~\ref{tab1})
imply that most of the BH-BH mergers occur within the first few Gyr of cluster evolution 
for both mergers within the cluster and mergers of escaped BH binaries.

The above results imply that an important class of candidates for dynamically
forming BH binaries that merge at the present epoch are star clusters
with initial mass $M_{cl} \ga 3\times 10^4 \Ms$\footnote{To correlate with the
observed clusters, we consider the mass $M_{cl}$ of the parent cluster,
\ie, the cluster mass before the BHs are formed through stellar evolution, 
which are heavier than the clusters of low mass stars that we model, for the
same value of $N$. For a given simulated cluster,
we estimate the corresponding parent mass $M_{cl}$ by weighting the mean stellar mass of
$\langle m \rangle_{cl}\approx 0.6 \Ms$ of a Kroupa IMF with star from 
$0.1\Ms$ upto $100\Ms$ with the total number of stars $N$ for that cluster.}, which are less than few Gyr old.
Such clusters represent intermediate-age massive clusters (hereafter IMC) with 
initial masses close to the upper-limit of the 
initial cluster mass function (ICMF) in spiral \citep{wkl2004,lrs2009a}
and starburst \citep{gls2006} galaxies.
While it is not impossible to obtain BH-BH mergers within
a Hubble time from lower-mass clusters, the overall BH-BH merger and escape
rates strongly decrease with cluster-mass as Table~\ref{tab1} indicates.
For $M_{cl} \la 1.5 \times 10^4 \Ms$, mergers already become much
rarer (see Table~\ref{tab1}). Due to the statistical nature of
merger or ejection events it is ambiguous to set any well-defined limit on the cluster-mass
beyond which these events become appreciable (such an estimate would also
require a much larger number of N-body integrations). In view of our results,
$M_{cl} \approx 3\times 10^4 \Ms$ is a representative lower limit beyond which an
appreciable number of mergers and escapers merging within a Hubble time can be obtained.

Old globular clusters, which can be about 10 times or more massive, are expected to produce
mergers or escapers more efficiently. As the timescale of depletion of BHs
from the BH cluster is nearly independent of the parent cluster mass (see Sec.~\ref{tdist}),
GCs can also be expected to produce BH-BH mergers over similar
time-span as the IMCs, \ie, within the first few Gyr of evolution.
Since GCs are typically much older ($\sim 10 {\rm ~Gyr}$),
they do not contribute significantly to the present-day merger rate, since most
of the mergers from them would have occurred earlier. Considering the
light-travel time of $\approx 4.5 {\rm ~Gyr}$ from the maximum distance
$D\approx 1500 {\rm ~Mpc}$ form which these BH-BH binaries can be detected by
``AdLIGO''(see below), only GCs close to the above distance
could contribute detectable events, mostly from escaped BH-BH binaries.

On the other hand, young massive clusters with ages less than 50 Myr, representing star clusters near
the high-mass end of the ICMF \citep{lrs2009b}, are generally too young
to produce BH-BH mergers. All models in Table~\ref{tab1} produce mergers
significantly later than this age (except one escaped BH binary in each of the
models C65K110 and C100K200). Hence, IMCs seem to be most likely
candidates for producing observable BH-BH mergers dynamically. 

\subsection{Detection rate}\label{rate}

We now make an estimate of the BH-BH merger detection rate from IMCs
by ground-based GW observatories like LIGO and AdLIGO. In estimating
the overall BH-BH merger rate using the results of our model clusters, one needs to consider the
distribution of the cluster parameters that are varied over the models, \viz,
cluster mass, half-mass radius, and BH retention fraction. Such distributions
are far from being well determined, except for the mass distribution for
young clusters in spiral and starburst galaxies \citep{bik2003,blm2003,gls2004}.
Therefore, determination of an overall merger rate considering the distribution
of our computed clusters can be ambiguous. Hence, as a useful alternative,
we determine the BH binary merger detection rates for each of the cluster models
in Table~\ref{tab1} that gives an appreciable number of mergers, for a representative
density of IMCs. Such an approach has been considered by earlier authors, \eg, 
\citet{olr2006} and can provide a reasonable idea of the rate of detection of
BH-BH mergers from IMCs.

As an estimate of the space density of IMCs, we adopt that for young populous clusters
in \citet{pzm2000}, which has a similar mass-range as the IMCs:
\begin{equation}
\rho_{cl} = 3.5 {\rm ~} h^3 {\rm ~Mpc}^{-3},
\label{eq:rypc}
\end{equation}
where $h$ is the Hubble parameter, defined as $H_0/100{\rm ~km~s}^{-1}$,
$H_0$ being the Hubble constant \citep{pbl93}. The above space density 
has been derived from the space densities
of spiral, blue elliptical and starburst galaxies \citep{hey97} assuming that
young populous clusters have the same specific frequencies ($S_{N}$) as old GCs
\citep{vdb95,mcl99}, but in absence of any firm determination
of the $S_{N}$s of the former. We compute the detection rate for each model
cluster assuming that it has a space density of the above value.

The LIGO/AdLIGO detection rate of BH-BH mergers from a particular model cluster can be
estimated from (\citealt{bky2007} and references therein)
\begin{equation} 
{\mathcal R}_{LIGO} = \frac{4}{3}\pi D^3\rho_{cl}{\mathcal R}_{mrg},
\label{eq:ligorate}
\end{equation}
where ${\mathcal R}_{mrg}$ is the compact binary merger rate from a cluster and
$D$ is the maximum distance from which the emitted GW from
a compact-binary inspiral can be detected. $D$ is given by
\begin{equation} 
D = D_0 \left(\frac{M_{ch}}{M_{ch,nsns}}\right)^{5/6},
\label{eq:range}
\end{equation}
where $D_0=18.4$ and 300 Mpc for LIGO and AdLIGO respectively.
The quantity $M_{ch}$ is the ``chirp mass'' of the compact binary
with component masses $m_1$ and $m_2$, which is given by
\begin{equation}
M_{ch} = \frac{(m_1 m_2)^{3/5}}{(m_1 + m_2)^{1/5}},
\label{eq:chirp}
\end{equation}
and $M_{ch,nsns}=1.2\Ms$ is that for a binary with two
$1.4\Ms$ neutron stars.

For a BH binary with $m_1=m_2=10\Ms$, $M_{ch}=8.71\Ms$ which gives
$D \approx 1500{\rm ~Mpc}$ for AdLIGO. The AdLIGO detection rates
${\mathcal R}_{\rm AdLIGO}$ (mean over 3 Gyr taking into account the
time of escape $t_{esc}$ of the escaped binaries) for the model clusters are shown in
Table~\ref{tab1}, where the currently accepted value of the Hubble parameter
$h=0.73$ is assumed. The error in each detection rate is simply obtained from
the Poisson dispersion of the total number of mergers for the corresponding
cluster. Note that these detection rates are for clusters with solar-like metallicity
which is implied by our assumption of $10\Ms$ BHs for all the clusters.

To obtain a basic estimate of the overall detection rate of BH-BH mergers from
IMCs, we consider the subset of our computed models that are isolated clusters with full BH retention
(see Table~\ref{tab1}). We take the
mass function of the IMCs to be a power law with index $\alpha = -2$ which is
the approximate index of the ICMF in spiral and starburst galaxies
(see \eg, \citealt{gls2004,lrs2009a}). Then the weighted average of the corresponding AdLIGO detection
rates is $\overline{\mathcal R}_{\rm AdLIGO} \approx 31(\pm 7)$ yr$^{-1}$, which estimates
the total present-day detection rate of BH-BH mergers from IMCs expected for AdLIGO.
The corresponding LIGO detection rate is negligible,
$\overline{\mathcal R}_{\rm LIGO} \approx 7.4\times 10^{-3}{\rm ~yr}^{-1}$.
Note that these BH-BH detection rates
are only lower limits. First, the observed population of star clusters can be
an underestimation by a factor of 2 owing to their dissolution in the
tidal field of their host galaxies (see \citealt{pzm2000} and references therein).
Second, the above detection rates are only
from IMCs and there can be additional contributions from GCs (see above).

In comparing the AdLIGO detection rates from our computations with those from earlier
works, we note that our rates are typically an order of magnitude smaller than
those of \citet{pzm2000}, but about one order of magnitude larger than those
obtained by \citet{olr2006}. The principal origin of the former difference is due to
the fact that while we considered only IMCs distinguishing them as the most appropriate
candidates for producing present-day BH-BH mergers (see above),
\citet{pzm2000} also included GCs, which have considerably larger spatial density 
and also larger fraction of BH-BH binaries merging within the Hubble time, as obtained
from their analytic extrapolations. Note that the number of escapers
as obtained by them for young populous clusters and the fraction of them merging within
a Hubble time (see Table~1 of \citealt{pzm2000}) is similar to that obtained from the
present computations, implying qualitative agreement. The above authors apparently
did not consider the time scales of formation and depletion of the BH sub-systems in
their preliminary study. On the other hand, although \citet{olr2006} considered clusters significantly
more massive than our's in their Monte-Carlo approach, so that larger merger detection
rates can be expected, their clusters were
much older (8 Gyr and 13 Gyr) than IMCs which, in accordance with our results,
accounts for their much lower detection rate.

It is interesting to note that the dynamical BH-BH merger detection rates obtained by us are typically
an order of magnitude higher than that from
primordial stellar binaries as predicted by \citet{bky2007}
based on their revised binary evolution model, and is similar to that for the isolated NS-NS
binaries derived by them. Hence our results imply that dynamical BH-BH binaries constitute
the dominant contribution to the BH-BH merger detection. Thus,
the dynamical BH-BH inspirals from star clusters seem to be a promising channel
for GW detection by the future AdLIGO, although their estimated detection rate with the
present LIGO detector is negligible, in agreement with the hitherto non-detection of GW.

\subsection{Limitations and outlook}\label{lim}

The work presented here is a first step towards a detailed study of
the dynamics of stellar-mass BHs in star clusters and the consequences
for GW-driven BH mergers, and improvements in several directions are possible.
First, we do not consider the initial phase of the cluster here when BHs
form through stellar evolution and a more consistent approach would be to
begin with a star cluster with a full stellar spectrum and produce
the BHs from stellar evolution. Note however that in the present study
we have inserted numbers of BHs in our clusters of low-mass stars
similar to what would have been formed from the 
stellar evolution of a cluster following a Kroupa IMF (see Sec.~\ref{runs}).
Stellar evolution also produces NSs (about twice in number as the BHs)
which also segregate to the central region of the cluster. It is interesting
to study the dynamics of the NS cluster and how it is affected by the
(more concentrated) BH cluster --- in particular the formation of tight
inspiralling NS-NS and NS-BH binaries, which are important for both GW detection
and GRBs.

Another aspect that we do not consider in our present study is
the effect of primordial stellar binaries. Tight stellar binaries aid the
formation of compact binaries through double-exchanges \citep{gr2006}, in
addition to the 3-body mechanism (see Sec.~\ref{intro}) which can
increase the number BH-BH (also BH-NS and NS-NS) binaries formed and hence
the merger rate. Thus the study of BH-BH binaries in star clusters with
primordial binaries is an important next step. Such studies are in progress and will be
presented in future papers. 

Finally, the number of N-body computations in this initial study is not enough
to obtain the BH-BH merger rate as a function of cluster mass
and BH retention fraction with a reasonable accuracy. There are typically one
integration per cluster model. To obtain merger rate dependencies
with cluster parameters, \eg, mass, half mass radius, binary fraction and BH retention,
many computations are needed within smaller intervals, which involves much larger time and computing capacity
than that utilized for the present project. Such results, combined with improved
knowledge of cluster parameter distributions and BH formation through supernova explosions of massive
stars, would provide a robust estimate of the BH-BH merger detection rate. Conversely,
with such a merger rate function, the detection of BH-BH mergers by GW detectors like AdLIGO in the near future
will shed light on the above mentioned long-standing questions.

\section*{Acknowledgments}

We are thankful to Sverre Aarseth of the Institute of Astronomy, Cambridge, UK, for
many discussions and suggestions and his timely modifications of the NBODY6 code, which have
been very helpful for our computations. We thank Keigo Nitadori for his
development and regular updates of the GPU libraries utilized in NBODY6.
We thank the anonymous referee for very helpful comments and suggestions which have
improved several parts of this paper significantly.
This work has been supported by the Alexander von Humboldt Foundation.


\label{lastpage}
\end{document}